\documentclass{ws-mpla}

\begin{document}

\markboth{Johan Bijnens and Joaquim Prades}
{The Hadronic Light-by-Light Contribution to the 
Muon $g-2$: Where Do We Stand?}

\catchline{}{}{}{}{}

\thispagestyle{empty}
\begin{flushright}
CERN-PH-TH/2007-035\\
LU TP 07-05\\
hep-ph/0702170
\end{flushright}
\vspace{2cm}
\begin{center}

{\large\bf The Hadronic Light-by-Light Contribution to the 
Muon Anomalous Magnetic Moment: Where Do We Stand?
}\\
\vfill
{\bf  Johan Bijnens$^{a)}$
 and Joaquim Prades$^{b)}$
\footnote{On leave of absence from CAFPE and Departamento de
 F\'{\i}sica Te\'orica y del Cosmos, Universidad de Granada, 
Campus de Fuente Nueva, E-18002 Granada, Spain.}
}\\[0.5cm]

$^{a)}$  Department of Theoretical Physics, Lund University\\
S\"olvegatan 14A, S-22362 Lund, Sweden.\\[0.5cm]

$^{b)}$ Theory Unit, Physics Department, CERN \\ 
CH-1211  Gen\`eve  23, Switzerland.\\[0.5cm]

\end{center}
\vfill
\begin{abstract}
\noindent
We review the status of
the hadronic light-by-light contribution to the muon anomalous 
magnetic moment
and critically compare recent calculations. We also study in detail which
momentum regions the $\pi^0$ exchange main contribution originates.
We  also argue that $a_\mu^{light-by-light} = (11\pm4) \times 10^{-10}$
 encompasses the present understanding of this contribution
and comment on some directions to  improve on that.
\end{abstract}
\vfill
February 2007
\newpage
\setcounter{page}{1}
\setcounter{footnote}{0}

\title{The Hadronic Light-by-Light Contribution to the 
Muon Anomalous Magnetic Moment: Where Do We Stand?
}

\author{\footnotesize JOHAN BIJNENS}

\address{Department of Theoretical Physics, Lund University,\\
S\"olvegatan 14A, SE 22362 Lund, Sweden\\
bijnens@thep.lu.se}

\author{JOAQUIM PRADES\footnote{On leave of absence from CAFPE and
Departamento de
F\'{\i}sica Te\'orica y del Cosmos, Universidad de Granada, 
Campus de Fuente Nueva, E-18002 Granada, Spain.}}

\address{Theory Unit, Physics Department, CERN \\ 
CH-1211  Gen\`eve  23, Switzerland\\
prades@ugr.es
}

\maketitle

\pub{Received (Day Month Year)}{Revised (Day Month Year)}

\begin{abstract}
We review the status of
the hadronic light-by-light contribution to the muon anomalous 
magnetic moment
and critically compare recent calculations. We also study in detail which
momentum regions the $\pi^0$ exchange main contribution originates.
We  also argue that $a_\mu^{light-by-light} = (11\pm4) \times 10^{-10}$
encompasses the present understanding of this contribution
and comment on some directions to  improve on that.

\keywords{Muon; Anomalous magnetic moment}
\end{abstract}

\ccode{PACS Nos.: 13.40.Em; 11.15.Pg; 12.20.Fv; 14.60.Ef}

\newcommand{\be}{\begin{equation}}
\newcommand{\ee}{\end{equation}}
\newcommand{\ba}{\begin{eqnarray}}
\newcommand{\ea}{\end{eqnarray}}
\newcommand{\dis}{\displaystyle}

\newcommand{\e}{e}

\section{Introduction}

The  muon anomalous magnetic moment
$g-2$ [$a_\mu \equiv (g-2)/2$]  has been measured by the 
E821 experiment (Muon g-2 Collaboration)
at Brookhaven National Laboratory (BNL)
 with an impressive accuracy of 0.72 ppm \cite{BNL06}
yielding the present
world average\cite{BNL06}
\be
a_\mu^{\rm exp} = 11 \, 659 \, 208.0(6.3) \times 10^{-10} \, 
\ee
with an accuracy of 0.54 ppm.
 New experiments\cite{ROB06,MRR06} are under design
with a goal of measuring $a_\mu$ with an accuracy of at least 0.25 ppm.

On the theory side,  a large amount of work
has been devoted to reduce the
uncertainty of the Standard Model prediction. A recent updated
discussion with an extensive list of references
for both theoretical predictions  and experimental results  
is Ref.~\refcite{MRR06} and Ref. \refcite{PAS05},
 and a more introductory exposition
can be found in the lectures by Knecht in Ref. \refcite{Knechtlectures}.

\begin{figure}[ht]
\label{lbl}
\begin{center}
\epsfig{file=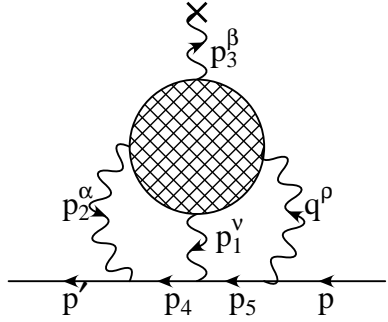,width=5cm}
\end{center}
\caption{The hadronic light-by-light
contribution to the muon $g-2$.}
\end{figure}

In this paper, we review the present status
of the hadronic light-by-light contribution (hLBL).
A somewhat shorter version is the published talk
in Ref. \refcite{kazimierz}.
The uncertainty in the hLBL is expected to
eventually become the largest theoretical error.
This contribution is shown schematically in Fig.~\ref{lbl}.
It consists of three photon legs coming from
the  muon line connected to
the external electromagnetic field
by hadronic processes.
Its contribution can be written as
\ba
\label{Mlbl}
\dis{\cal M}
&=& \vert e\vert^7 V_\beta
\int \frac{{\rm d}^4 p_1}{ (2\pi )^4}
\int \frac{{\rm d}^4 p_2}{ (2\pi )^4} \,
\frac{1}{ q^2\, p_1^2 \, p_2^2 (p_4^2-m^2) \,
(p_5^2 - m^2)} 
\nonumber \\
&\times&
  \Pi^{\rho\nu\alpha\beta} (p_1,p_2,p_3) \,
\bar{u}(p^\prime )\gamma_\alpha (\not{\! p}_4 +m )\gamma_\nu
(\not{\! p}_5 +m ) \gamma_\rho u(p) \, 
\ea
where $q=p_1+p_2+p_3$. 
To obtain the amplitude ${\cal M}$ in (\ref{Mlbl}),
the hadronic contribution to the
full correlator $\Pi^{\rho\nu\alpha\beta}(p_1,p_2,p_3 \to 0)$
needs to be known for all possible four-momenta $p_1$ and $p_2$.
The correlator is defined via
\ba
\label{four}
\Pi^{\rho\nu\alpha\beta}(p_1,p_2,p_3)&=&
i^3 \int {\rm d}^4 x \int {\rm d}^4 y
\int {\rm d}^4z \e^{i (p_1 \cdot x + p_2 \cdot y + p_3 \cdot z)}
\times \nonumber \\
&&
 \langle 0 | T \left[  V^\rho(0) V^\nu(x) V^\alpha(y) V^\beta(z)
\right] |0\rangle 
\ea
with $V^\mu(x)=\left[ \overline q \hat Q \gamma^\mu q \right](x)$
and $\hat Q = {\rm diag}(2,-1,-1)/3$ the quark charges.
The external magnetic field couples to the photon leg with momentum
 $p_3 \to 0 $.
In the remainder whenever we refer to $a_\mu$ we specifically mean only
the hadronic light-by-light contribution to it.

Clearly, the correlator (\ref{four}) is a complicated object.
It contains many independent Lorentz structures, each of comes
with a function of the variables $p_1^2$, $p_2^2$ and $q^2$.
As a consequence,
many different energy scales can be involved in the
calculation of the  
hadronic light-by-light contribution to muon $g-2$.
This makes it difficult to obtain the full needed behavior
of the correlator (\ref{four}) from known constraints.
Therefore no full first principles
calculation exists at present.
The needed results cannot be directly
related to measurable quantities either. 
Lattice QCD calculations are at the exploratory stage only, see e.g.
Ref.~\refcite{HBIY06}.

In fact, there has long been a confusion about hadronic exchanges\footnote{We
stick here to the formulation ``exchange'' as used by us\cite{BPP96}.
 It is often referred to as
``pole'' contributions. We consider this misleading because the exchanged
particle is used off-shell.}
versus quark loop estimates. This confusion
was resolved by organizing the different contributions
according to the lowest power in $1/N_c$ and the lowest order
in the chiral perturbation theory (CHPT)
 expansion counting where they start contributing\cite{EdR94}.
One can distinguish four types of contributions:
\begin{itemize}
\item Goldstone boson exchange contributions are order $N_c$ and start
contributing at order $p^6$ in CHPT.
\item (Constituent) quark-loop 
and non-Goldstone boson exchange contributions  are order
$N_c$ and start contributing at order $p^8$ in CHPT.
\item Goldstone boson loop contributions are order  one in $1/N_c$
and start contributing at order $p^4$ in CHPT.
\item Non-Goldstone boson loop contributions are order one in $1/N_c$
 and start to contribute at order $p^8$ in CHPT.
\end{itemize}

The two existing {\em full} calculations\cite{BPP96,HK98},
are based on this classification.
The Goldstone boson exchange contribution (GBE) was shown to be numerically
dominant in Refs. \refcite{BPP96} and \refcite{HK98} 
after strong cancellations between the other 
contributions. But the other contributions, though each
smaller than the GBE, were not separately negligible.
Using effective field theory techniques,
Ref.~\refcite{KNPR02} showed that the leading double logarithm comes
from the GBE and was positive. 
Refs.~\refcite{KNPR02} and \refcite{KN02} found a global sign mistake
in the GBE of the earlier work\cite{BPP96,HK98} which was confirmed by
the authors of those works\cite{BPP01,HK01} and by others\cite{BCM02,RW02}.
In the remainder we will always correct for this sign mistake without
explicitly mentioning it.

Recently, Melnikov and Vainshtein pointed out new short-distance
constraints on the correlator (\ref{four}) 
in Ref.~\refcite{MV04}, 
studied and extended in Ref.~\refcite{KPPR04}. 
The authors of Ref.~\refcite{MV04} constructed a model
which satisfies their main new short-distance
constraints in order to study its effects and found a number
significantly different from the earlier work.
They approximated the full hLBL
by the GBE and axial-vector
exchange contributions. 

One of the purposes of this review is to critically compare the different
contributions in the different calculations and extend somewhat
on our earlier comments\cite{kazimierz}.
For the dominant GBE we
also present some new results on the momentum regions which are relevant.
In earlier work several studies were done to check which momentum
regions were important. These used different methods, varying the vector
meson mass\cite{HK98}, studying the cut-off dependence\cite{BPP96}
and expansions around various momentum regions in the loop
integrals\cite{MV04,Melnikov}.

In Sect.~\ref{pre2002} we discuss the calculations done before
2002 and compare their results. Sect.~\ref{after2002} discusses the short
distance constraints proposed by Melnikov and Vainshtein and the numerical
results presented in their paper.
In Sect.~\ref{GBE} we show in detail in which momentum regions
the contributions from $\pi^0$ exchange originate 
and Sect.~\ref{Comparison} compares and comments on the
various contributions in the different calculations.
Finally, we present our conclusions as to the present best value
and error for the hLBL in Sect.~\ref{conclusions}.

\section{Results obtained up to 2002}
\label{pre2002}

In this section we discuss the calculations performed in 
the period 1995-2001. These were organized according to 
the large $N_c$ and CHPT countings \cite{EdR94} discussed above.
The CHPT counting is used as a classification tool, none of these
calculations were actually performed at a fixed order in CHPT.
We want to emphasize once more that the calculations in
Refs. \refcite{BPP96,HK98,BPP01} and \refcite{HK01} 
showed that only after several large
cancellations in the rest of the contributions,
the numerically dominant one is the Goldstone boson exchange.
In this section we concentrate on the work in Refs. \refcite{BPP96}
and \refcite{BPP01},
with some comments and results from Refs. \refcite{HK98,KN02}
 and \refcite{HK01}.

\subsection{Pseudo-Scalar Exchange}

\begin{figure}
\begin{center}
\epsfig{file=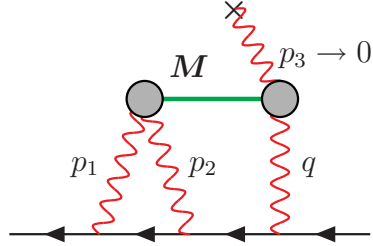}
\end{center}
\caption{A generic meson exchange contribution to the hadronic 
light-by-light
part of the muon $g-2$.}
\label{figexchange}
\end{figure}

The pseudo-scalar exchange was saturated
by the Goldstone boson exchange in Refs. \refcite{BPP96,HK98,BPP01}
and \refcite{HK01}.
This contribution is shown in Fig.~\ref{figexchange}
with $M=\pi^0,\eta,\eta^\prime$.

Refs \refcite{BPP96} and \refcite{BPP01},
used a variety of $\pi^0 \gamma^* \gamma^*$ form factors
\be
{\cal F}^{\mu\nu} (p_1,p_2) 
\equiv {N_c}/({6 \pi}) \,({\alpha}/{f_\pi})
\, i \, \varepsilon^{\mu\nu\alpha\beta} p_{1\alpha}
p_{2 \beta} \, {\cal F}(p_1^2, p_2^2)
\ee
fulfilling several possible QCD constraints.
A more extensive analysis of this form factor
was done in Ref. \refcite{BP01} finding very similar
numerical results.
In particular, the three-point form factors
${\cal F} (p_1^2,p_2^2)$ used in Refs.~\refcite{BPP96} and
\refcite{BPP01}
had the correct QCD short-distance behavior\footnote{
The observance of QCD short-distance constraints was implemented
for this one and several other contributions in 
Refs.~\refcite{BPP96} and \refcite{BPP01},
contrary to the often heard wrong claim that Ref.~\refcite{MV04}
is  the first calculation to take such constraints into account,
e.g. see Ref. \refcite{ET06}.}
\ba
\label{pi0OPE}
{\cal F} (Q^2,Q^2)  &\to& {A}/{Q^2} \, ,\quad
{\cal F} (Q^2,0)  \to {B}/{Q^2} \, ,
\ea
when $Q^2$ is Euclidean. These form factors were in agreement
with available data including the
slope at the origin as well as treating the  $\pi^0$, $\eta$ and $\eta'$
mixing.
All form factors converged for a cutoff scale 
$\Lambda \sim (2 - 4)$ GeV and produced  small  numerical differences
when plugged into the hadronic light-by-light contribution.

Somewhat different ${\cal F} (p_1^2,p_2^2)$ form factors
where used in Refs. \refcite{HK98,KN02} and \refcite{HK01} 
but the results agree well. For comparison, one can find 
the results of Refs. \refcite{BPP96,HK98,KN02,BPP01}
and \refcite{HK01} in Tab.~\ref{tab1} 
of the $\pi^0$ exchange and
 after  adding $\eta$ and $\eta'$ exchange contributions to the
dominant $\pi^0$ one.
\begin{table}
\begin{center}
\tbl{Results for the $\pi^0$ and 
$\pi^0$, $\eta$ and $\eta'$ exchange contributions.
\label{tab1}}
{
\begin{tabular}{c|cc}
 &\multicolumn{2}{c}{ $10^{10} \times a_\mu$}\\
 & $\pi^0$ only &  $\pi^0$, $\eta$ and $\eta'$\\
\hline
Bijnens, Pallante and Prades \cite{BPP96,BPP01} & 5.6  & 8.5 $\pm$ 1.3 \\
Hayakawa and  Kinoshita \cite{HK98,HK01}  & 5.7 & 8.3 $\pm$ 0.6 \\
Knecht and Nyffeler \cite{KN02}
($h_2=0$) & 5.8 & 8.3 $\pm$ 1.2\\
Knecht and Nyffeler \cite{KN02} ($h_2=-10$~GeV$^2$) & 6.3 & \\
Melnikov and Vainshtein \cite{MV04} &  7.65 &11.4$\pm$1.0
\end{tabular}}
\end{center}
\end{table}

\subsection{Axial-Vector Exchange}

This contribution is depicted in Fig.~\ref{figexchange}
with $M=A=a_1^0,f_1$ and possibly other axial-vector resonances.
For this contribution one needs the $A\gamma\gamma^*$ and
$A\gamma^*\gamma^*$ form factors.
Little is known about these but there exist anomalous Ward identities
which relate them to the $P\gamma\gamma^*$ and $P\gamma^*\gamma^*$
form factors.

This contribution was not studied by Knecht and Nyffeler\cite{KN02}.
Refs. \refcite{BPP96,HK98,BPP01} ands \refcite{HK01} used nonet symmetry, 
which is exact in the large
$N_c$ limit, for the masses of the axial-vector resonances.
Their results are shown
in Tab.~\ref{tab2} for comparison.
\begin{table}
\begin{center}
\tbl{Results for the axial-vector exchange contributions.
\label{tab2}}{
\begin{tabular}{c|c}
 Axial-Vector Exchange Contributions
 & $10^{10} \times a_\mu$\\
\hline
Bijnens, Pallante and Prades \cite{BPP96,BPP01}  & 0.25 $\pm$ 0.10 \\
Hayakawa and  Kinoshita \cite{HK98,HK01}  & 0.17 $\pm$ 0.10\\ 
Melnikov and Vainshtein \cite{MV04} & 2.2$\pm$0.5
\end{tabular}}
\end{center}
\end{table}

\subsection{Scalar Exchange}

This contribution is shown in Fig.~\ref{figexchange}
with $M=S=a_0,f_0$ and possible other scalar resonances.
For this contribution one needs the $S\gamma\gamma^*$ and
$S\gamma^*\gamma^*$ form factors. 
Within the extended Nambu--Jona-Lasinio (ENJL)
 model used in Refs.~\refcite{BPP96} and \refcite{BPP01}, chiral 
Ward identities impose relations between the constituent
quark loop and scalar exchanges. The needed scalar form factors
are also constrained at low energies by CHPT.
Refs. \refcite{BPP96} and \refcite{BPP01} 
used nonet symmetry for the masses.
This contribution was not included by the other 
groups\cite{HK98,HK01,MV04}.

The leading logarithms of the scalar exchange are the same as 
those of the pion exchange but with opposite sign\cite{BCM02}.
Refs.  \refcite{BPP96} and \refcite{BPP01} 
find that sign for the full scalar exchange
contribution, obtaining
\be
a_\mu ({\rm Scalar}) = - (0.7\pm0.2) \times 10^{-10} \, .
\ee

\subsection{Other contributions at leading order in $1/N_c$.}

\begin{figure}
\begin{center}
\epsfig{file=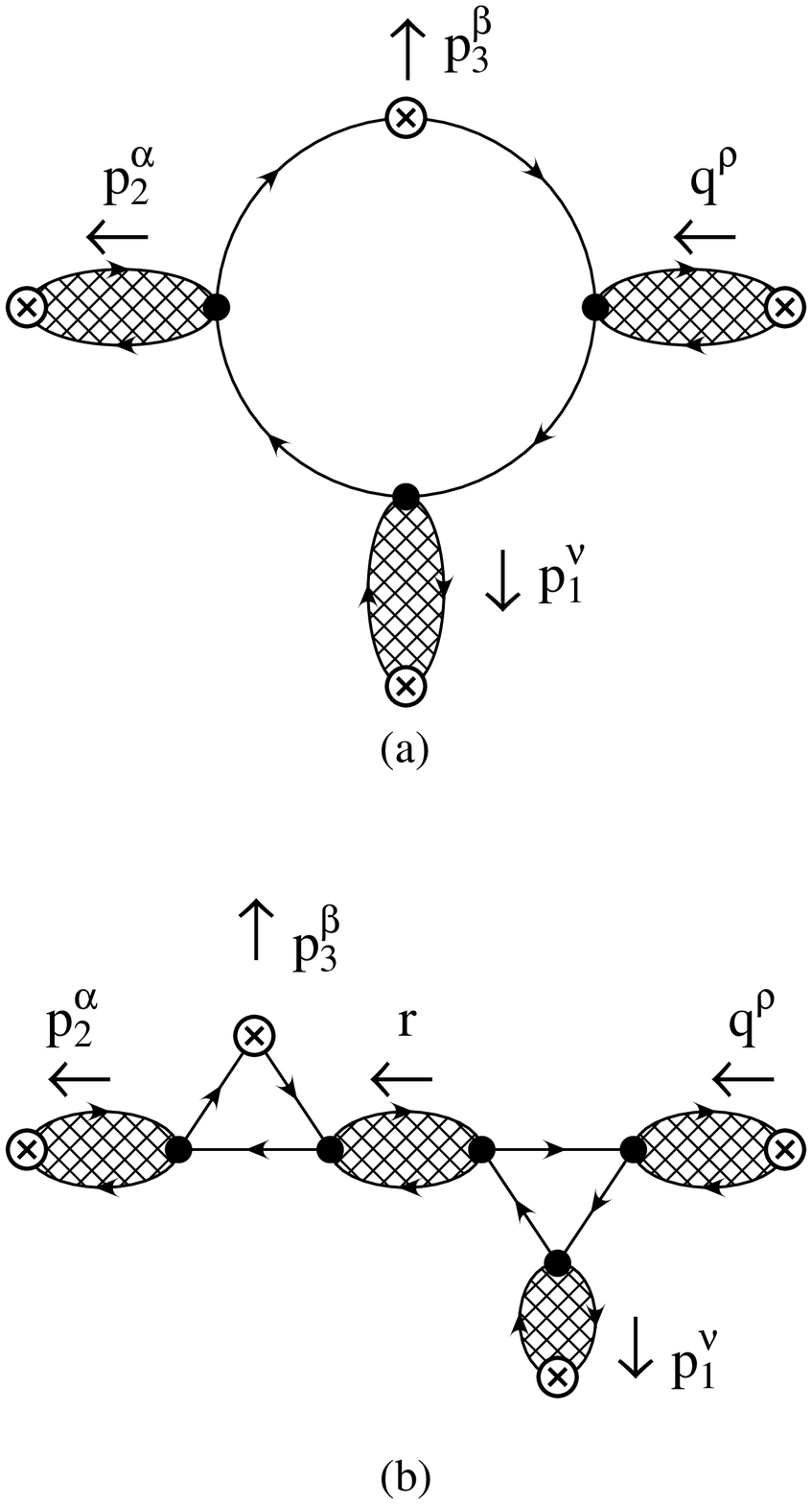,width=5cm,clip}
\end{center}
\caption{Quark-loop  contribution, as modeled in ENJL.}
\label{figquarkloop}
\end{figure}

This includes any contributions that are not modeled by
exchanged particles. At short-distance, the main one is the quark-loop.
At long distances they are often modeled as a constituent quark-loop
with form factors in the couplings to photons. 
This corresponds to the contribution 
shown in Fig.~\ref{figquarkloop}.
Refs.~\refcite{BPP96} and \refcite{BPP01} 
split up the quark momentum integration into two pieces
 by introducing an Euclidean matching scale $\Lambda$. 
At energies
below $\Lambda$, the ENJL model was used
to compute the quark-loop contribution  while above $\Lambda$
 a bare (partonic) heavy  quark loop of mass
$\Lambda$ was used. The latter part scales as $1/\Lambda^2$ and
mimics the high energy behavior of QCD for a massless quark with
an IR cut-off  around $\Lambda$ --see footnote $^b$.
  Adding these two contributions yields
a stable result 
as can be seen in Tab.~\ref{quarkL}. 
\begin{table}
\begin{center}
\tbl{Sum of the short- and long-distance quark loop contributions\protect\cite{BPP96}
as a function of the matching scale $\Lambda$.
\label{quarkL}}{
\begin{tabular}{c|cccc}
$\Lambda$ [GeV] & 0.7 & 1.0 & 2.0 &4.0\\
\hline
\rule{0cm}{13pt} $10^{10} \times a_\mu$ & 2.2 &  2.0& 1.9& 2.0
\end{tabular}}
\end{center}
\end{table}

\subsection{NLO in $1/N_c$: Goldstone Boson Loops}

At next-to-leading order (NLO) 
in $1/N_c$, the leading contribution in the chiral counting 
to the correlator
in (\ref{Mlbl}), corresponds to charged pion and Kaon loops
which can be depicted analogously 
 to the quark-loop in Fig.~\ref{figquarkloop} 
but with charged pions and Kaons running
inside the loop instead. In general one expects loops of heavier particles
to be suppressed and has only been
evaluated for the pion loop and the much smaller Kaon loop.

The needed form factors\footnote{Note that neither the ENJL model nor
any fixed order in CHPT
was used in any of the estimates of this contribution.}
$\gamma^*  P^+P^-$ and  $\gamma^* \gamma^* P^+P^-$  vertices
were studied extensively in Ref. \refcite{BPP96}.
 In particular which form factors were fully
compatible with chiral Ward identities were studied.
The full vector
meson dominance model (VMD) is one model fulfilling the known constraints.
The conclusion unfortunately is that there is a large ambiguity in the
momentum dependence starting at order $p^6$ in CHPT. Both 
the full VMD model\cite{BPP96,BPP01} and the hidden gauge symmetry (HGS)
model\cite{HK98,HK01} satisfy the known constraints.
Unfortunately, this ambiguity cannot easily be resolved
since there is no data for $\gamma^* \gamma^* \to \pi^+ \pi^-$.
Adding the charged pion and Kaon
loops, the results obtained in Refs. \refcite{BPP96}
and \refcite{HK98} 
are listed in Tab.~\ref{tab3}.
\begin{table}
\begin{center}
\tbl{ Results for the charged and Kaon loop contributions
to the hadronic light-by-light contribution to muon $g-2$.
\label{tab3}}{
\begin{tabular}{c|c}
 Charged Pion and Kaon Loop Contributions
 & $10^{10} \times a_\mu$\\
\hline
Bijnens, Pallante and Prades (Full VMD)
\cite{BPP96,BPP01}  & $-$1.9 $\pm$ 0.5 \\
Hayakawa and  Kinoshita (HGS) \cite{HK98,HK01}  & $-$0.45 $\pm$ 0.85\\
 Melnikov and Vainshtein (Full NLO in $1/N_c$ guess) \cite{MV04} & 0$\pm$1 
\end{tabular}}
\end{center}
\end{table}

In view of this model dependence, the authors of Refs.
\refcite{BPP96} and \refcite{BPP01} considered that
the difference between the results from Ref. \refcite{BPP96} and 
Ref. \refcite{HK98}
for this contribution needs to be added
 {\em linearly} to the  final  uncertainty of the hadronic
light-by-light contribution to $a_\mu$. 

\section{New Short-Distance Constraints}
\label{after2002}

Melnikov and Vainshtein pointed out\cite{MV04} a new
short-distance constraint on the correlator (\ref{four}).
This constraint is for
\ba
\label{OPEMV}
\langle T [ V^\nu(p_1) V^\alpha(p_2) V^\rho(-q=-p_1-p_2) ]| \gamma(p_3 \to 0)
\rangle
\ea
and follows from the OPE for two vector currents
when $P_1^2\simeq P_2^2 \gg Q^2$ with $P_1^2=-p_1^2$, $P_2^2=-p_2^2$
and $Q^2=-q^2$:
\ba
\label{OPEVV}
T[ V^\nu(p_1) V^\alpha(p_2) ] \sim 
\varepsilon^{\nu\alpha \mu\beta} \, (\hat p_\mu/\hat p^2) \, 
[\overline q \hat Q^2 \gamma_\beta \gamma_5 q] (p_1+p_2)
\ea
with $\hat p =  (p_1-p_2)/2 \simeq p_1 \simeq -p_2$
and $\hat Q$ is the light quark electrical charge matrix
 (\ref{four}). This constraint was afterward
generalized in Ref. \refcite{KPPR04}. Note that the new part is the
use of (\ref{OPEVV}) for the full correlator (\ref{four}).
Short-distance
 was already used to obtain the first constraint in (\ref{pi0OPE}).

The authors of Ref.~\refcite{MV04} saturated the full correlator
by exchanges. The new OPE constraint
is satisfied by introducing a pseudo-scalar exchange 
with the vertex on the $q,p_3$ side of Fig.~\ref{figexchange} point-like
rather than including a form factor.
This change strongly breaks the symmetry between the two ends
of the exchanged particle. There are also  OPE constraints for 
$P_1^2\approx P_2^2\approx Q^2$ and 
$P_2^2\approx Q^2 \gg P_1^2$,
 essentially derived from the quark-loop behavior in this 
regime\cite{MV04}.  Both latter OPE constraints on the
correlator (\ref{four})
are not satisfied by the model used in Ref. \refcite{MV04}
but they argued that this made only a small numerical 
difference of order $0.05 \times 10^{-10}$.

Ref. \refcite{MV04} added to the pseudo-scalar exchange an axial-vector
exchange contribution.
They found this contribution to be extremely sensitive
to the mixing of the resonances
 $f_1(1285)$ and $f_1(1420)$  as can be seen
in Tab.~\ref{massmixing}, taken from the results there.
\begin{table}
\begin{center}
\tbl{Results quoted in Ref. \protect\refcite{MV04} for
the pseudo-vector exchange depending of the $f_1(1285)$ 
and $f_1(1420)$ resonances mass mixing.
\label{massmixing}}{
\begin{tabular}{c|c}
Mass Mixing  & $10^{10} \times a_\mu$\\
\hline
No OPE and Nonet Symmetry with M=1.3 GeV& 0.3  \\
New OPE  and Nonet Symmetry with M= 1.3 GeV &  0.7   \\
New OPE  and Nonet Symmetry with M= M$_\rho$ &  2.8   \\
New OPE  and Ideal Mixing with Experimental Masses &  2.2 $\pm$ 0.5\\   
\end{tabular}}
\end{center}
\end{table}
The difference between the lines labeled ``No OPE'' and ``New OPE''
is the effect of making the $q,p_3$ vertex point-like.
The authors of Ref. \refcite{MV04} took the ideal mixing result for their
final result for $a_\mu$.

\section{Momentum Regions for $\pi^0$ Exchange}
\label{GBE}

We were somewhat puzzled by the effect when saturating the
 new short distance constraint by GBE in Ref. \refcite{MV04} 
and have therefore done a few studies to see whether
the changes there 
come from large momentum regimes or are located elsewhere.
This was because our total estimate of the
quark-loop was similar to the numerical 
change in the GBE of Ref. \refcite{MV04}.
In order to do this study, we have adapted the method used
in Refs. \refcite{BPP96} and \refcite{BP01} 
to various form factors used in earlier works. 
We rotate the integrals in (\ref{Mlbl}) into Euclidean space.
The eight dimensional integral can be easily reduced to a five dimensional
integral. Here one can choose as variables\footnote{In Refs.
 \refcite{BPP96} and \refcite{BP01}
 a different set was used not quite as suitable
for the present study.}
 $P_1$, $P_2$ and $Q$
and two angles $\theta_1$ and $\theta_2$. These are the angles between
the Euclidean $p_1$, $p_2$ and the muon momentum while 
$P_1$, $P_2$ and $Q$ are the size of the Euclidean momenta
with  $P_1^2=-p_1^2$, $P_2^2=-p_2^2$ and $Q^2=-q^2$.
Ref.~\refcite{KN02} performed the integrals over three of these quantities
analytically but not for the asymmetric case used by Ref.~\refcite{MV04}.
We have therefore use numerical integration.
The main integration routine used by us earlier\cite{BPP96,BP01} was VEGAS.
For the present study we have also performed the integration using an
adaptive Gaussian multidimensional integration routine and have checked
for several quantities that both agree and reproduce earlier known results.

We will show the contributions to the muon anomalous magnetic moment
from $\pi^0$ exchange for several different form factors.
These correspond to the point-like $\pi^0\gamma^*\gamma^*$
form factor (WZW), the full vector meson dominance model (VMD),
the LMD+V form factor\cite{KN02} with $h_2=-10$~GeV$^2$ (KN)
and the latter form factor but with the point-like
version on the soft-photon end\cite{MV04} (MV).
We will refer to these form factors
as WZW, VMD, KN, and MV in the remainder of this
section. We have used the values $h_1=0, h_5 = 6.93$~GeV$^2$ and
the value of $h_7$ as given by Ref. \refcite{KN02}. We picked the value
of $h_2$ that was argued\cite{MV04} to better produce subleading OPE
constraints. It raises the central value somewhat compared to
$h_2=0$ as shown in Tab.~\ref{tab1}

As inputs we used $M_V=M_{V_1}=0.770$~GeV
and $M_{V_2}= 1.465$~GeV, $F_\pi=92.4$~MeV and the measured $\pi^0$ 
and muon masses. This is the origin of the minor differences with
Ref.\refcite{KN02}.

As a first indication where the contributions to $a_\mu$ come from, we have
listed in Tab.~\ref{tab5} the value of $a_\mu$ for the four cases
with the constraint $Q,P_1,P_2 < \Lambda$.
\begin{table}
\tbl{$\pi^0$-exchange  results for $10^{10} \times a_\mu$ 
with a cut-off on the three photon momenta
for the four cases described in the text.
The last column is the difference between MV and KN form factors.
The numerical error is at or below the last digit quoted.
\label{tab5}}
{\begin{tabular}{c|ccccc}
Cut-off $\Lambda$ (GeV) & WZW & VMD & KN & MV & MV$-$KN\\
\hline
0.5 &  4.74& 3.37 & 3.39 & 3.68 & 0.29\\
0.7 &  7.51& 4.41 & 4.47 & 5.01 & 0.54\\
1.0 &  11.3& 5.14 & 5.29 & 6.15 & 0.86\\
2.0 &  21.9& 5.60 & 5.99 & 7.34 & 1.35\\
4.0 &  33.8& 5.65 & 6.20 & 7.79 & 1.59\\
8.0 &  49.6& 5.65 & 6.24 & 7.92 & 1.69\\
16.0&  68. & 5.64 & 6.23 & 7.96 & 1.73
\end{tabular}
}
\end{table}
We have shown the logarithmically square
divergent point-like case here to show
the size of the suppression introduced by the form factors.
Note that we cannot reproduce the 7.65 of Ref.~\refcite{MV04} but we do
reproduce the results of Refs. \refcite{BPP96,KN02} and \refcite{BP01}.
The new short-distance constraint (\ref{OPEVV}) came from the region
$Q\ll P_1\approx P_2$. We have thus checked how much of the difference
and total comes from the region with $Q < {\rm min}(P_1,P_2)$ 
and from the region with $Q$ larger than at least one of $(P_1$,$P_2)$, 
the  numbers quoted are for
$\Lambda = 16$ GeV. The numbers are $10^{10} \times a_\mu$.
\be
\mbox{
\begin{tabular}{cccc}
$Q<\min(P_1,P2)$: &  4.01 (\mbox{KN}) & 4.74 (\mbox{MV}) 
& 0.73 (\mbox{MV}$-$\mbox{KN}) \\ 
$Q>\min(P_1,P_2)$:&  2.24 (\mbox{KN}) & 3.23 (\mbox{MV}) 
& 0.99 (\mbox{MV}$-$\mbox{KN}) 
\end{tabular}}
\ee
As one sees, in fact, most of the difference comes from the region
where the OPE condition is strongly violated.

\begin{figure}
\begin{center}
\epsfig{file=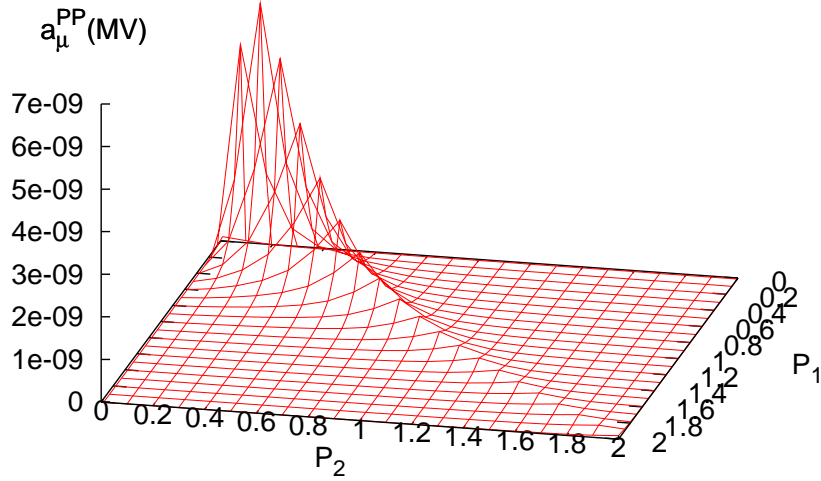,width=0.9\textwidth}
\end{center}
\caption{The quantity $a_\mu^{PP}$ of Eq.~\ref{defda})
as a function of $P_1$ and $P_2$
for the MV choice.
\label{figmvPP}}
\end{figure}

\begin{figure}
\begin{center}
\epsfig{file=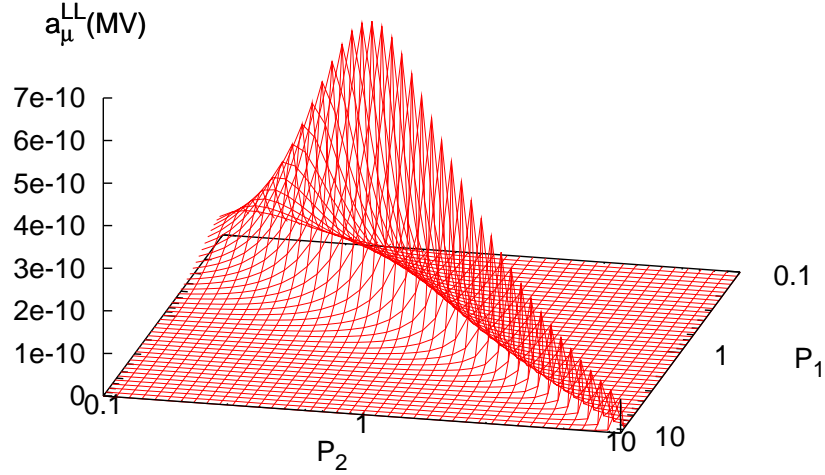,width=0.9\textwidth}
\end{center}
\caption{The quantity $a_\mu^{LL}$ of Eq.~\ref{defda})
as a function of $P_1$ and $P_2$
for the MV choice. $a_\mu$ is directly related to the
volume under the surface as plotted.
\label{figmvLL}}
\end{figure}

\begin{figure}
\begin{center}
\epsfig{file=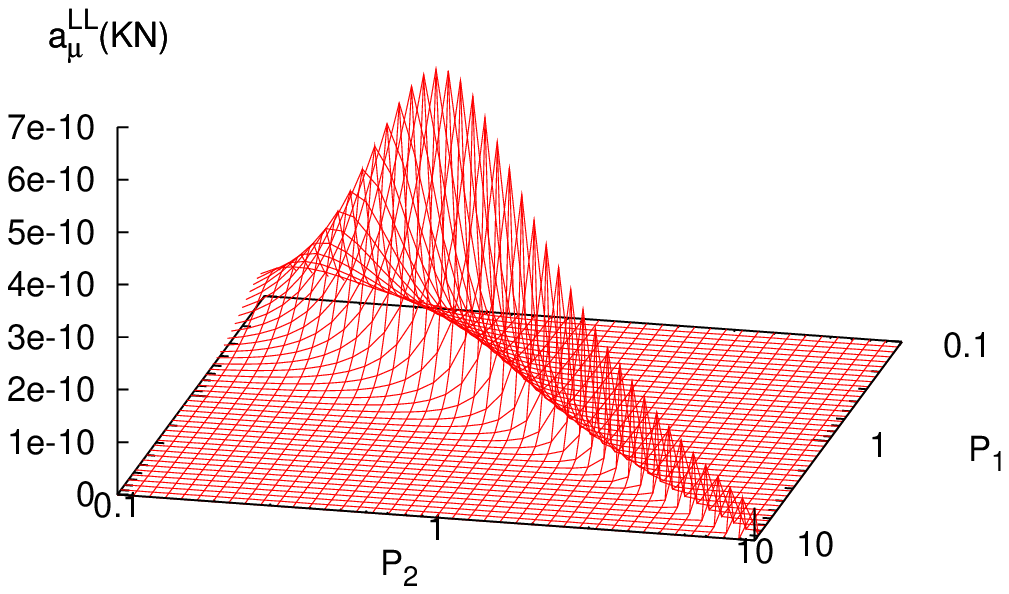,width=0.9\textwidth}
\end{center}
\caption{The quantity $a_\mu^{LL}$ of Eq.~\ref{defda})
as a function of $P_1$ and $P_2$
for the KN choice. $a_\mu$ is directly related to the
volume under the surface as plotted.
\label{figknLL}}
\end{figure}

\begin{figure}
\begin{center}
\epsfig{file=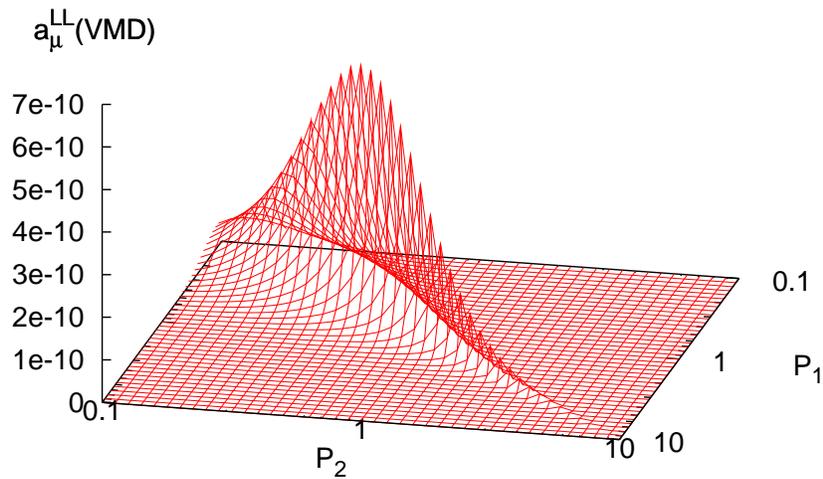,width=0.9\textwidth}
\end{center}
\caption{The quantity $a_\mu^{LL}$ of Eq.~\ref{defda})
as a function of $P_1$ and $P_2$
for the VMD choice. $a_\mu$ is directly related to the
volume under the surface as plotted.
\label{figvmdLL}}
\end{figure}

The results in Tab.~\ref{tab5} give only a partial indication 
of which momentum
regions are important. In the remaining figures we therefore show
the contribution to $a_\mu$ in several ways.
We always denote $p_1,p_2$ as the momenta on the $\pi^0$ side with both
photons connected to the muon line and $q$ the momentum on the 
soft-photon side.
We can thus rewrite the contribution to $a_\mu$ of (\ref{Mlbl}) in various
ways:
\ba
\label{defda}
a_\mu &=& \int dP_1 dP_2\,\, a_\mu^{PP}(P_1,P_2)
\nonumber\\
&=& \int dl_1 dl_2\,\, a_{\mu}^{LL}(l_1,l_2)
\nonumber\\
&=& \int dl_1 dl_2 dl_q\,\, a_{\mu}^{LLQ}(l_1,l_2,l_q) \, ,
\nonumber\\
{\rm with} \quad l_1&=& \log(P_1/\mbox{GeV}),\quad
l_2 = \log(P_2/\mbox{GeV}), \quad {\rm and} \quad 
l_q =  \log(Q/\mbox{GeV}) \, .
\ea
In Fig.~\ref{figmvPP} we have plotted $10^{10} \times a_\mu^{PP}(P_1,P_2)$
as a function of $P_1$ and $P_2$.
In this way of plotting it is however rather difficult 
to see why the contribution
with at least one scale above 1~GeV is as large 
as shown in Tab.~\ref{tab5}.
The quantity $a_\mu^{LL}$ is much more suitable for this.
The result for $a_\mu$ after integrating for this quantity is directly
proportional to the volume under the surface as it is plotted in 
Figs.~\ref{figmvLL}, \ref{figknLL} and \ref{figvmdLL} with a 
logarithmic scale
for $P_1$ and $P_2$. We have used the same scale for all three plots.
What one finds is that the VMD one has much smaller contributions
for $P_1$ and $P_2$ large but both MV and KN show a significant
contribution even at fairly high values of $(P_1,P_2)$. 
Also the contribution
at these higher values of $(P_1,P_2)$ is concentrated along the
axis $P_1=P_2$. One also see by comparing Figs.~\ref{figmvLL} 
and \ref{figknLL}
that the enhancement of the MV result over the KN result comes not from
a very different shape but more a general increase over the entire region.
The parts below $0.1$~GeV were not plotted, these are very similar for
all three cases. A plot for the WZW case simply shows a constantly growing
ridge along $P_1=P_2$ which produces then the $\log^2\Lambda$ divergence.

\begin{figure}
\begin{center}
\epsfig{file=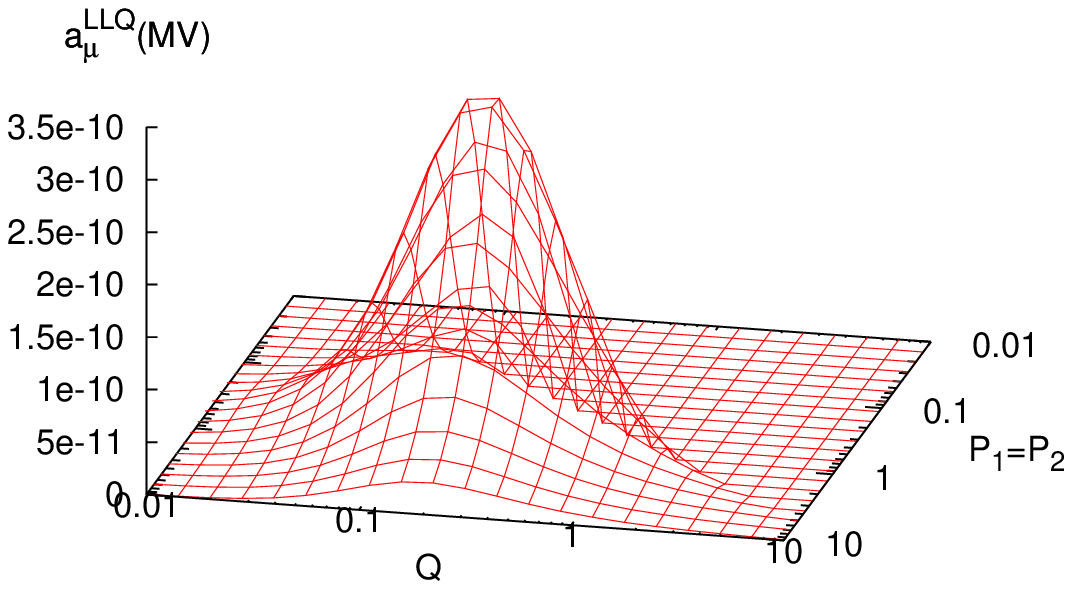,width=0.9\textwidth}
\end{center}
\caption{The quantity $a_\mu^{LLQ}$ of Eq.~\ref{defda})
as a function of $Q$ and $P_1=P_2$
for the MV choice. $a_\mu$ is directly related to the
volume under the surface as plotted.
\label{figmvLLQ}}
\end{figure}

\begin{figure}
\begin{center}
\epsfig{file=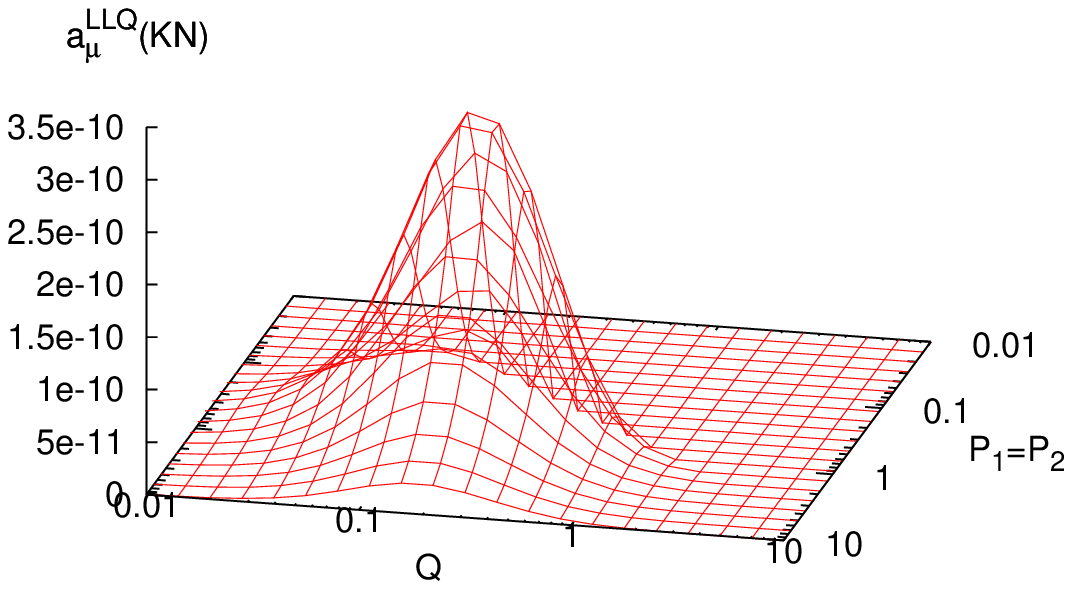,width=0.9\textwidth}
\end{center}
\caption{The quantity $a_\mu^{LLQ}$ of Eq.~\ref{defda})
as a function of $Q$ and $P_1=P_2$
for the KN choice. $a_\mu$ is directly related to the
volume under the surface as plotted.
\label{figknLLQ}}
\end{figure}

The figures before give an indication of which ranges of $(P_1,P_2)$ 
are important. But what about the values of $Q$ that are relevant.
This will of course depend on the values of $P_1$ and $P_2$.
We show in Fig.~\ref{figmvLLQ} the value for $a_\mu^{LLQ}$
along the line $P_1=P_2$. Again, the contribution to $a_\mu$ is
proportional to the volume under the surface as shown.
This is shown for the MV and KN form factors in Figs.~\ref{figmvLLQ} and
\ref{figknLLQ} respectively.
One surprise for us was that while one can see that the tail towards
larger values of $Q$ is somewhat larger for the MV form factor 
than the KN one, 
it is much less than expected and only marginally visible in the plot.
The main conclusion from this section is that the numerical difference
between MV and KN comes from relatively low values of $Q$ and moderate
values of $P_1$ and $P_2$. We have provided plots and numerics so that
readers can draw their own conclusions.

\section{Comparison}
\label{Comparison}

Let us now try to compare the different results of
the three calculations in Refs. \refcite{BPP96,BPP01,HK98,HK01}
and \refcite{MV04}.
In Tab.~\ref{comparisontab}, the results 
to leading order in $1/N_c$ are shown.
The quark loop is of the same order and has to be {\em added}
to get the full hadronic light-by-light while the model used
in Ref. \refcite{MV04} is saturated just by exchanges.
\begin{table}
\begin{center}
\tbl{Full hadronic light-by-light contribution
to $a_\mu$ at ${\cal O}(N_c)$. The difference between the
two results of Refs. \protect\refcite{BPP96} and \protect\refcite{BPP01}
 is the contribution of the
scalar exchange $-(0.7\pm0.1) \times 10^{-10}$.
This contribution is not included in Refs.
\protect\refcite{HK98,HK01} and  \protect\refcite{MV04}.
\label{comparisontab}
\label{largeN}}{
\begin{tabular}{c|c}
Hadronic light-by-light 
at ${\cal O} (N_c)$  & $10^{10} \times a_\mu$\\
\hline
Nonet Symmetry + Scalar Exchange \protect\cite{BPP96,BPP01} & 
10.2 $\pm$ 1.9\\ 
Nonet Symmetry \protect\cite{BPP96,BPP01}&  10.9 $\pm$ 1.9  \\
Nonet Symmetry \protect\cite{HK98,HK01} &  9.4 $\pm$ 1.6 \\ 
 New OPE and 
Nonet Symmetry \protect\cite{MV04} &  12.1 $\pm$ 1.0 \\ 
 New OPE and 
Ideal Mixing  \protect\cite{MV04} &  13.6 $\pm$ 1.5  
\end{tabular}}
\end{center}
\end{table}
In the GBE the effect of the new OPE in Ref. \refcite{MV04}
is a little larger than the quark loop
contributions of Refs. \refcite{BPP96} and \refcite{BPP01}
 but compatible within one sigma.
 This contribution has been discussed
in more detail in the previous section.
The new OPE in Ref. \refcite{MV04} 
similarly increases the axial-vector exchange
with nonet symmetry from 0.3 $\times 10^{-10}$ 
to 0.7 $\times 10^{-10}$
One thus sees a reasonable agreement in the comparison
of the  ${\cal O}(N_c)$
results of Refs. \refcite{BPP96,BPP01,HK98,HK01} and \refcite{MV04}
when using the  same mass mixing for the axial-vectors, namely, 
(10.9$\pm$ 1.9, 9.4 $\pm$ 1.6,  12.1 $\pm 1.0$). 

The final differences are due to the additional increase of
1.5$\times 10^{-10}$ from the ideal mixing in the axial vector exchange
in Ref. \refcite{MV04} and the scalar exchange of $-$0.7$\times 10^{-10}$
in Refs. \refcite{BPP96} and \refcite{BPP01}.
 
Let us now see what the different predictions at NLO in $1/N_c$ are.
In Ref. \refcite{MV04}, the authors studied the chiral expansion
of the charged pion  loop using the HGS model used in
Refs. \refcite{HK98} and \refcite{HK01}. 
This model is known not to give the correct QCD high energy
behavior in some two-point functions, in particular it does not fulfill 
Weinberg Sum Rules, see e.g. Ref. \refcite{BPP96}.
Within this model, Ref. \refcite{MV04} showed that there is 
a large cancellation between the first three terms 
of  an expansion of the charged pion loop contribution
in powers of $(m_\pi/M_\rho)^2$.
It is not clear how one should interpret this. In Refs. 
\refcite{BPP96}
some studies of the cut-off dependence of this contribution were done
and the bulk of their final number came from fairly low energies
which should be less model dependent.
However, it is clear that there is a large model dependence 
in the NLO in $1/N_c$ contributions.
But simply taking it to be
$(0\pm 1) \times 10^{-10}$
as  in Ref. \refcite{MV04} is rather drastic and certainly has
an underestimated error. 
The argument of very large higher order
corrections when expanded in CHPT orders which was used against this
contribution in Ref. \refcite{MV04} also applies to the $\pi^0$ exchange as
can be seen from Tab.~\ref{tab5} by comparing the WZW 
column with the others.

Let us now compare the results for the full hadronic 
light-by-light contribution to $a_\mu$ when summing all contributions.
The final result quoted in Refs. \refcite{BPP96,BPP01}, 
\refcite{HK98,HK01} and \refcite{MV04} 
can be found  in Tab.~\ref{final}. The apparent agreement
between Refs. \refcite{BPP96,BPP01} and \refcite{HK98,HK01} final number
is hiding
non-negligible differences which numerically compensate to a large extent.
There are differences in the quark loop and charged
pion and Kaon loops and Refs. \refcite{HK98,HK01} do not 
include the scalar exchange.
\begin{table}
\begin{center}
\tbl{Results for the full hadronic light-by-light contribution
to $a_\mu$.
\label{final}}{
\begin{tabular}{c|c}
Full Hadronic Light-by-Light & $10^{10} \times a_\mu$\\
\hline
Bijnens, Pallante and Prades
\cite{BPP96,BPP01}&  8.3  $\pm$ 3.2\\ 
Hayakawa and Kinoshita \cite{HK98,HK01}&  8.9  $\pm$ 1.7\\ 
Melnikov and Vainshtein \cite{MV04}&  13.6  $\pm$ 2.5\\ 
\end{tabular}}
\end{center}
\end{table}

Comparing the results of Refs. \refcite{BPP96,BPP01} and \refcite{MV04},
we have seen several differences of order
$1.5 \times 10^{-10}$, differences  which are not related
to the one induced by the new short-distance constraint
introduced in Ref. \refcite{MV04}.
These differences are numerically of the same order or smaller than 
the uncertainty quoted in Refs. \refcite{BPP96,BPP01} but tend to add up
making the total difference large as follows:
The different axial-vector mass mixing account for $-1.5 \times 10^{-10}$,
the absence of scalar exchange in Ref. \refcite{MV04} accounts
for $-0.7 \times 10^{-10}$ and the absence of the NLO in $1/N_c$
charged pion and Kaon loops contribution
 in Ref. \refcite{MV04} accounts for
$-1.9 \times 10^{-10}$. These model dependent
differences add up to $-4.1 \times 10^{-10}$ 
out of the final $-5.3 \times 10^{-10}$
difference between the results in Refs. \refcite{BPP96,BPP01}
and \refcite{MV04}.  
In addition we have shown from which regions in momentum
the main contribution originates.
Clearly, the new  OPE constraint
found in Ref. \refcite{MV04} alone does not account for the large 
final numerical difference with respect to
 Refs. \refcite{BPP96,BPP01} as a reading of it  seems to suggest.

\section{Conclusions}
\label{conclusions}

 At present, the only possible conclusion is that the 
situation of the hadronic
light-by-light contribution to $a_\mu$ is unsatisfactory.
However, looking into the various calculations
one finds a {\em numerical} agreement within roughly one sigma
when comparing the ${\cal O}(N_c)$ 
results found in Refs. \refcite{BPP96,HK98,BPP01,HK01} 
and \refcite{MV04},  see Tab.~\ref{largeN}. 
A new full ${\cal O}(N_c)$ calculation
studying the full correlator with the large
$N_c$ techniques developed in Refs. \refcite{BGLP03}
and \refcite{CEEKPP06}
and references therein, seems feasible and definitely desirable. 

At NLO in $1/N_c$, one needs to control  
both Goldstone and non-Goldstone boson loop contributions.
The high model dependence of the Goldstone boson loop is clearly visible
in the different results of Refs. \refcite{BPP96,BPP01}
and \refcite{HK98,HK01} and discussed in Refs. \refcite{BPP96} 
and \refcite{MV04}.
For non-Goldstone boson loops, little is known on how to 
consistently treat them, a recent attempt in another context is
Ref. \refcite{RSP04}.

In the meanwhile, we propose as an educated guess
for the total hLBL\footnote{
This educated guess agrees with the one
presented by Eduardo de Rafael \cite{MRR06}
and ourselves \cite{kazimierz} at the
 ``Final Euridice Meeting'' in Kazimierz, August 2006
and  by one of us (JB) at the ``DESY Theory Workshop'' in Hamburg,
September 2005.}
\ba
\label{finalpluserror}
a_\mu= (11  \pm 4) \times 10^{-10} \, .
\ea
We believe that, that this number and error
capture our present understanding of the hLBL
contribution to $a_\mu$.
This number can be reached using several different arguments:
the new short-distance constraint found in Ref. \refcite{MV04}
and the ideal mixing for the
axial-vector exchange should lead to some increase of the results
of Refs. \refcite{BPP96,BPP01} and \refcite{HK98,HK01}; 
the scalar exchange and the pion and Kaon loops
are expected to lead to some decrease of the result of Ref. \refcite{MV04};
one can also average the leading in $1/N_c$ results (three middle results
of Tab.~\ref{largeN}) which turn out to be within one sigma.
 The final error remains a guess but the
error in (\ref{finalpluserror}) is chosen to include all the known
uncertainties.

\section*{Acknowledgments}

This work is supported in part by the European Commission (EC) RTN network,
Contract No.  MRTN-CT-2006-035482  (FLAVIAnet), 
the European Community-Research Infrastructure
Activity Contract No. RII3-CT-2004-506078 (HadronPhysics) (JB),
the Swedish Research Council (JB), 
 MEC (Spain) and  FEDER (EC) Grant No.
FPA2006-05294 (JP), and Junta
de Andaluc\'{\i}a Grant Nos. P05-FQM-101  and P05-FQM-437 (JP).

\end{document}